\documentclass[a4paper,11pt]{article}
\pdfoutput=1 

\usepackage{jinstpub} 

\title{\boldmath SiW ECAL for future $e^+e^-$ collider}

\author[a,1]{V.~Balagura,\note{Corresponding author.}}
\author[b]{S.~Bilokin,}
\author[b]{J.~Bonis,}
\author[a]{V.~Boudry,}
\author[a]{J.-C.~Brient,}
\author[c]{S.~Callier,}
\author[g]{T.~Cheng,}
\author[a,d]{R.~Cornat,}
\author[c]{C.~De~La~Taille,}
\author[h]{T.H.~Doan,}
\author[a]{M.~Frotin,}
\author[a]{F.~Gastaldi,}
\author[f]{H.~Hirai,}
\author[j]{S.~Jain,}
\author[h]{Sh.~Jain,}
\author[d]{D.~Lacour,}
\author[d]{L.~Lavergne,}
\author[e]{A.~Lleres,}
\author[a]{F.~Magniette,}
\author[k]{L.~Mastrolorenzo,}
\author[a]{J.~Nanni,}
\author[b]{R.~Poeschl,}
\author[h]{A.~Pozdnyakov,}
\author[i]{A.~Psallidas,}
\author[g]{M.~Ruan,}
\author[a]{M.~Rubio-Roy,}
\author[c]{N.~Seguin-Moreau,}
\author[a]{K.~Shpak,}
\author[f]{T.~Suehara,}
\author[b]{A.~Thiebault,}
\author[k]{J.~Wright,}
\author[a,h]{D.~Yu}

\affiliation[a]{LLR - Ecole polytechnique, Palaiseau, France}
\affiliation[b]{LAL, Orsay, France}
\affiliation[c]{Omega - Ecole polytechnique, Palaiseau, France}
\affiliation[d]{LPNHE, Paris, France}
\affiliation[e]{LPSC, Grenoble, France}
\affiliation[f]{Kyushu University, Fukuoka, Japan}
\affiliation[g]{IHEP, Bejing, China}
\affiliation[h]{NCU, Taiwan}
\affiliation[i]{NTU, Taiwan}
\affiliation[j]{Tata Inst. of Fundamental Research, Mumbai, India}
\affiliation[k]{Imperial College, London, United Kingdom}

\emailAdd{balagura@llr.in2p3.fr}

\abstract{Calorimeters with silicon detectors have many unique features and
  are proposed for several world-leading experiments. We discuss the tests of
  the first three 18$\times18$~cm$^2$ layers segmented into 1024 pixels of the
  technological prototype of the silicon-tungsten electromagnetic calorimeter
  for a future $e^+e^-$ collider.  The tests have beem performed in November
  2015 at CERN SPS beam line.}


\proceeding{Instrumentation for Colliding Beam Physics, INSTR17\\
  27 February - 3 March 2017\\
  Novosibirsk, Russia}

\begin{document}
\maketitle
\flushbottom

\section{Silicon calorimeters in ILD and beyond}
\label{sec:intro}

Silicon detectors without amplification have excellent characteristics not
only for tracking devices but also for calorimeters. They are easily
segmentable, perfectly linear, independent of environmental changes and stable
in time.  The response to minimum-ionizing particle (MIP) is determined only
by the thickness of p--n junction and the gain of front-end electronics,
therefore it is uniform across the channels within a few percent and easy to
calibrate. Overall, with the silicon calorimeters one can achieve the best
granularity and the lowest systematics, which makes them ideal for a precise
electromagnetic calorimetry optimized for Particle Flow Algorithms (PFA) at
International Linear Collider (ILC)~\cite{ilc} or other future $e^+e^-$
collider. In addition, the silicon detectors are radiation hard which is of
utmost importance for Large Hadron Collider (LHC).  They can provide the
timing at the level of 20--50~psec which is also important for both LHC and
ILC.

The main disadvantage of the silicon detectors is a high cost. For large scale
production, like for International Large Detector (ILD)~\cite{ildsid}, the
expected price from Hamamatsu HPK company, Japan is about 2.5~EUR/cm$^2$. The
silicon calorimeter also requires a low-noise electronics. The intrinsic
stochastic energy resolution of the ILD silicon-tungsten electromagnetic
calorimeter (SiW ECAL) is 17--20\%/$\sqrt{E/\text{GeV}}$ depending on the
number of layers and the silicon thickness. This resolution is sufficient for
PFA, since the jet energy resolution is dominated by the hadronic calorimeter
measurement at low energies (below 100~GeV) and by the shower overlaps causing
``confusion'' in assigning calorimeter hits to the right particles at high
energies.

Silicon ECAL has been proposed for all future $e^+e^-$ collider experiments
(ILD, SiD~\cite{ildsid}, CEPC~\cite{cepc1,cepc2}, FCC~\cite{fcc},
CLIC~\cite{clic}).  It has been approved by CMS collaboration for the phase-II
upgrade of their calorimeter endcaps (High-Granularity CALorimeter,
HGCAL~\cite{hgcal}). It is planned to install 40 radiation-hard silicon layers
in each endcap.  Currently, HGCAL collaboration carries out an intensive
research on the shower timing at the level of $\sim$20~psec. It should allow
to localize the region of a primary interaction and to reduce effectively the
CMS phase-II pile-up of 140--200. The same principle is studied by ATLAS. They
propose High Granularity Timing Detector (HGTD)~\cite{hgtd}, a preshower in
front of the currently installed liquid argon calorimeter. Here, the smaller
signal, either MIP-like or from the very beginning of the shower, is amplified
in the silicon by a factor of $\sim10$ (``Low Gain Avalanche
Detector''). Finally, a few silicon layers with high granularity and fast
timing are proposed to be integrated in LHCb ECAL in phase II
upgrade~\cite{lhcb}. They should help to fight against the pile-up and also to
measure better the angle between $\pi^0$ photons and in this way to improve
its mass resolution.

\section{Tests of technological prototype at CERN SPS}

This paper describes the recent progress in SiW ECAL R\&D for ILD.  In
2004--2011 the so-called ``physical'' prototype of SiW ECAL with 30 layers has
been build and successfully tested within CALICE
collaboration~\cite{phys1}. It was designed to prove the physical PFA
principles in beam tests. The electronics was not embedded inside the
calorimeter but put aside.

After the physical prototype, the emphasis has moved to the technological
realization of the detector scalable to ILD ECAL. This second generation
prototype is called ``technological''. The electronics is embedded in the
active layer.  It is based on a dedicated 64 channel front-end chip
SKIROC~\cite{skiroc}. Each channel has 15 memory slots called SCA digitized by
12 bit Analog-to-Digital-Converter (ADC) linear in the range of about 1--1500
MIPs. SKIROC has auto-trigger capabilities as in ILD there is no central
trigger, the event is built using bunch crossing (BX) time stamps.

In the current design, ILC collides bunch trains during 1 msec and then stays
idle for 199 msec. The latter period is used to readout the data stored in
SKIROC. To save the power and to simplify the cooling, the front-end
electronics is switched off after each readout cycle. This is called "power
pulsing". In such a mode one SKIROC channel consumes only about 27~$\mu W$. In
practice, however, the electronics should be switched on before the bunch
trains so that the transition processes (causing e.g. pedestal drifts) finish
well in advance.

A printed circuit board (PCB) of the current prototype has 16 SKIROC chips and
1024 channels. It serves 4 silicon sensors each segmented into 256 pixels of
5.5$\times$5.5~mm$^2$ size. The pixels are glued to the corresponding PCB pads
with the conductive epoxy. This gluing technology was proven by several years
of tests of the physical prototype.

The technological prototype has already reached the density of channels
required for ILD.  The next steps will be to connect several such detector
elements in-line and on both sides of the absorber layer to form an ILD
detector element called ``slab''.

Two series of technological prototype tests have been performed at CERN SPS,
in November 2015 and in June 2016. By the time of the first test 4 layers each
with 1024 channels have been built and brought to the beam, 3 of them were
successfully operated during the tests. The SPS beam time was kindly provided
by CMS HGCAL. The second beam campaign was provided by Semi-Digital HCAL
(SDHCAL)~\cite{sdhcal} and was envisaged as a combined test of high
granularity ECAL and HCAL. By that time 10 ECAL layers have been built, 7 were
operational but 6 of them suffered from high noises. Only one layer was fully
operational.  Therefore, in the following we discuss only the SPS tests in
November 2015 with 3 layers. The next beam tests with standalone SiW ECAL are
planned in June 2017 in DESY, Germany.

\subsection{Pedestals}

In the current version of SKIROC chip there is no pedestal suppression: every
trigger initiates readout of pedestals in all non-triggered channels.  This
provides a lot of pedestal data which was useful for the debugging as the
determination of the pedestal positions was found to be not so trivial.

First, they should be calculated not only per channel but also per SCA memory
slot. With 15 memories for each channel, the total number of pedestals in one
layer is 1024$\times$15 = 15360. Second, there is a small fraction of events
where one or more channels trigger on negative signals possibly caused by a
pick-up noise from digital lines. The pedestals in other channels are
distorted so that such events should be excluded.

Finally, there is one more complication. The triggered event in one bunch
crossing (BX) may be followed by several "retriggers" in BX+1, BX+2, BX+3,
\ldots{} .  With low trigger threshold or in high noise conditions similar
retrigger sequences may appear spontaneously and may even dominate the data
stream.  This actually happened in June 2016 beam tests in 6 layers, when only
one was working properly at the nominal trigger threshold.

In November 2015 the retriggers were acceptable in all 3 readout layers.
However, in the events followed by retriggers the pedestals were shifted from
the right values. The MIP signal position was not affected, however. The
difference between triggered and pedestal data might arise from the fact that
the pedestals are recorded with some delay, namely, at the edge of the next
bunch clock.

\begin{figure}[htbp]
\centering
\includegraphics[width=7.61cm]{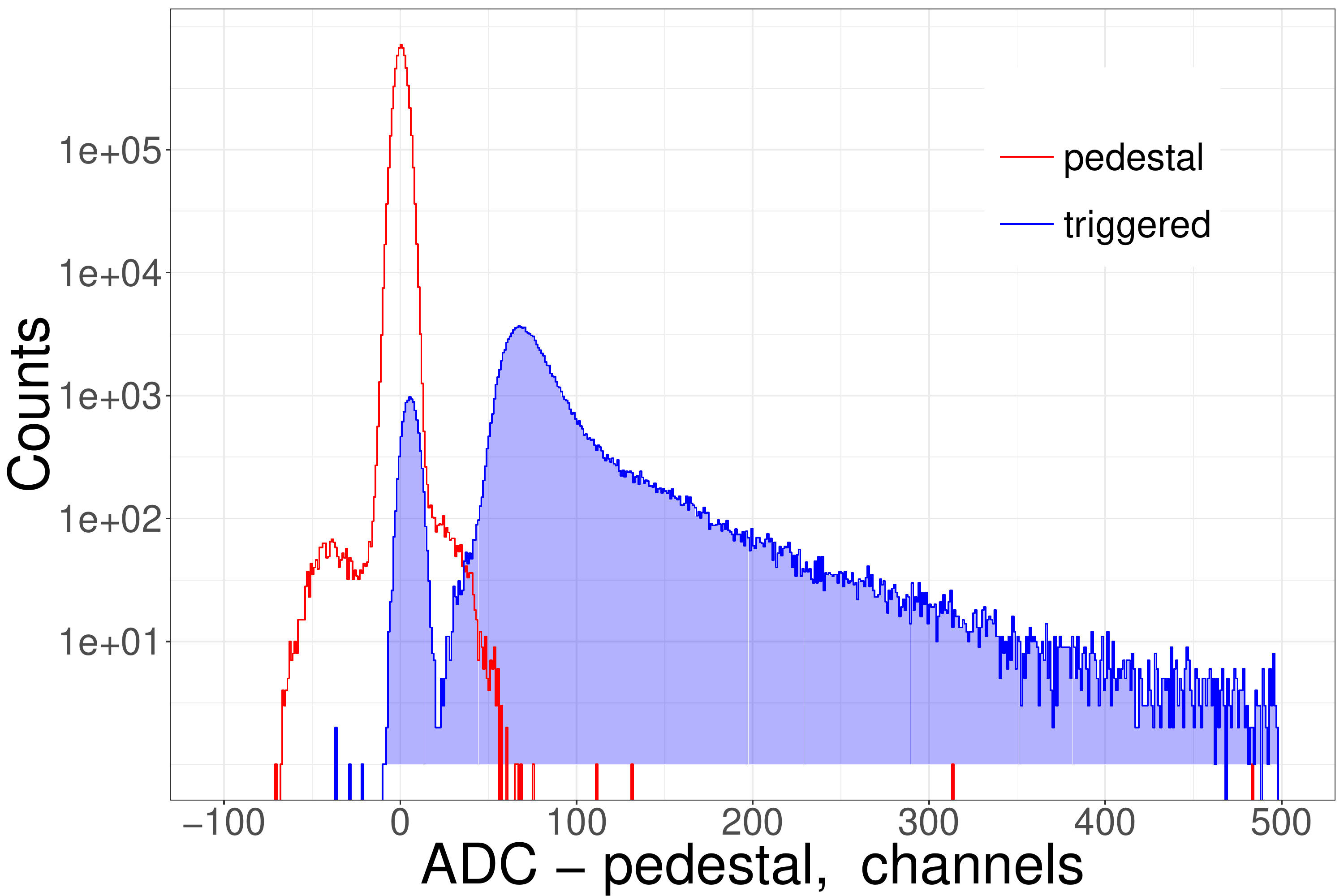}
\includegraphics[width=7.31cm]{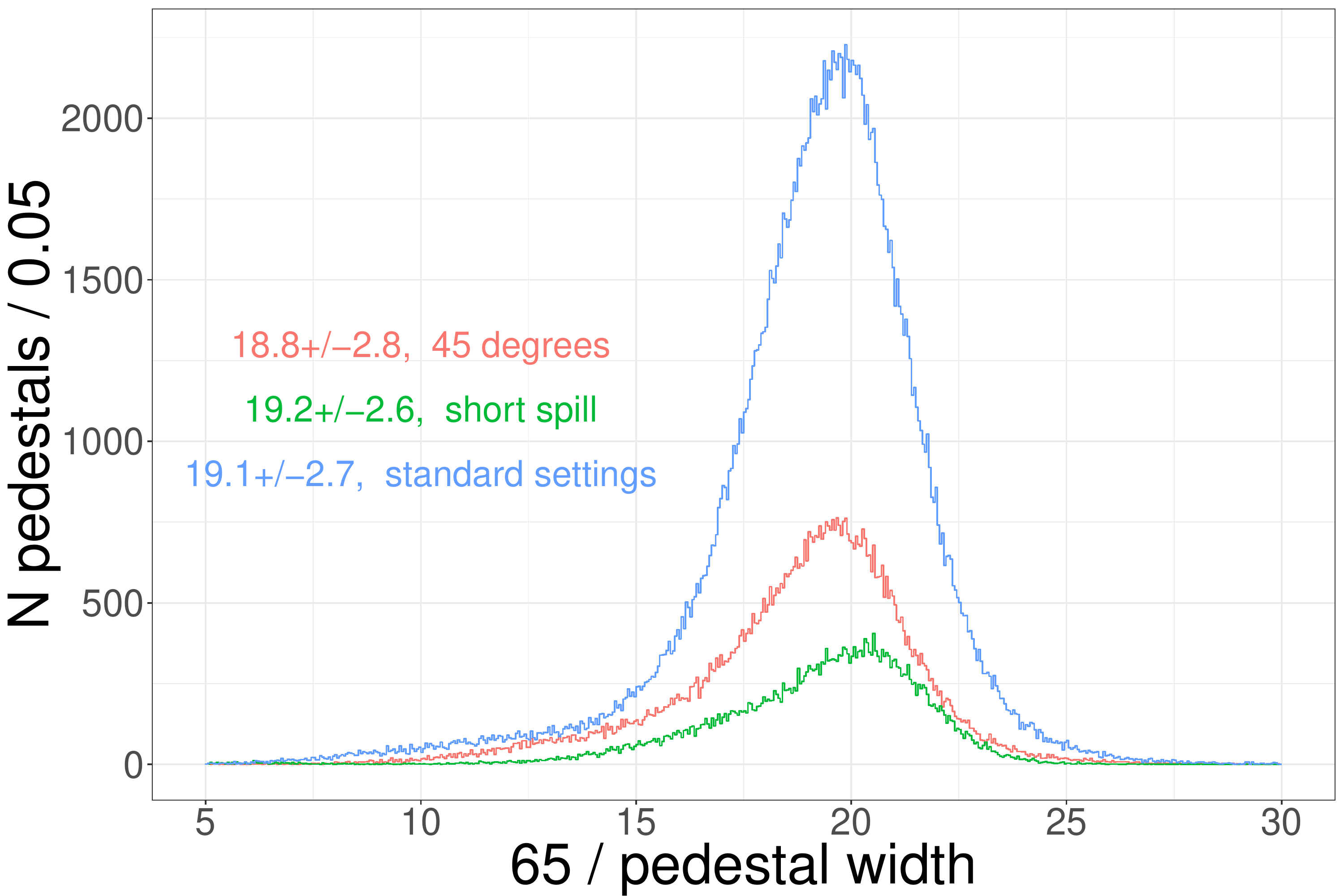}
\caption{\label{fig1}Left: the combined ADC spectra for all channels of one
  layer during two hours muon run. Triggered and pedestal data are shown
  separately. A peak at zero for the former is produced by spontaneous
  retriggers.  Right: the ratio of the MIP signal (65 channels) to the
  pedestal width for various running conditions.}
\end{figure}

In the end, the pedestals were calculated per channel, per SCA memory in the
events without any negative signals and not followed by retriggers.  A few
problematic channels (\#37 in all chips and \#45-47 in chips 1,9 with double
peak pedestals) are excluded.  The resulting pedestal spectra are
approximately Gaussian, as shown by red open histogram in Figure~\ref{fig1}
left for about 1000 channels in one layer. There is an excellent separation of
the pedestals and the MIP signals peaking at $\sim$65.

SKIROC gain is regulated by the configurable preamplifier feedback
capacitance. For an ILD dynamic range it should be set to the maximal value of
6~pF. It was found, however, that about twice better signal-over-noise ratio
can be obtained with five times higher gain corresponding to 1.2~pF. This
value was chosen for November 2015 beam tests. The fact that the noise
increases less than the gain at 1.2~pF means that the preamplifier noise is not
dominating in SKIROC.

Pedestal width reflecting the noises is studied in 19 two-hours muon runs
taken in 5 consecutive days.
The ratio of the MIP most-probable-value (65 ADC channels) to the standard
deviation of the pedestal spectrum per channel, SCA and muon run is shown in
Figure~\ref{fig1} right. It summarizes 316 000 pedestal spectra with at least
100 events accumulated in about 3000 channels.  Different colors distinguish
different running conditions: with 200~msec data taking period and with ECAL
perpendicular to the beam (``standard settings'') or turned later by about
45$^\circ$, and with 1.1~msec period to emulate ILC bunch trains (``short
spills''). One can see that in average signal-over-noise ratio is about 19,
which demonstrates an excellent potential of the SKIROC chip.

The measured pedestal positions stay constant in 18 muon runs with respect to
the first one (taken with the ``standard settings'') with a standard deviation
equal to 0.4\% of the MIP signal. There is one exception, however, when the
first layer was operated with 1.1~msec data taking period its pedestals
deviated in average by 3.8\% of the MIP. This indicates a pedestal drift
continuing beyond currently set 1.4~msec delay after switching on the
electronics in power pulsing mode. In other two layers no pedestal drift was
observed.

\subsection{Muon detection efficiency and gain uniformity}
Three ECAL layers are sufficient for an efficiency determination: one layer
can be studied when two others are put in coincidence and used as an ``offline
trigger''. 150~GeV muons have been used for this analysis.  Since they
traversed ECAL almost perpendicularly, for the trigger coincidence the hit
pixel position is required to be the same in both layers. For the layer under
study this requirement is relaxed so that the fired pixel can deviate by at
most one pixel in both coordinates. The time coincidence between the layers is
detected if the bunch crossing (BX) clock numbers (with the period of 400
nsec), differ by at most 1. Possible retriggers in BX+1, BX+2, \ldots{}
following the muon in BX are removed. The pixel is defined to be hit if it has
a SKIROC trigger flag set and its pedestal subtracted ADC value exceeds 20 (40
for the trigger). To suppress showers from 150 GeV muons it is required that
every trigger layer has only one hit.

There is one more complication: the inefficiency in the central chips with
high occupancy could be due to a limited SKIROC memory.  To exclude this
possibility it is required that the chip whose efficiency is studied, has at
least one empty memory slot in the end of data taking.

\begin{figure}[htbp]
\centering
\includegraphics[width=14cm]{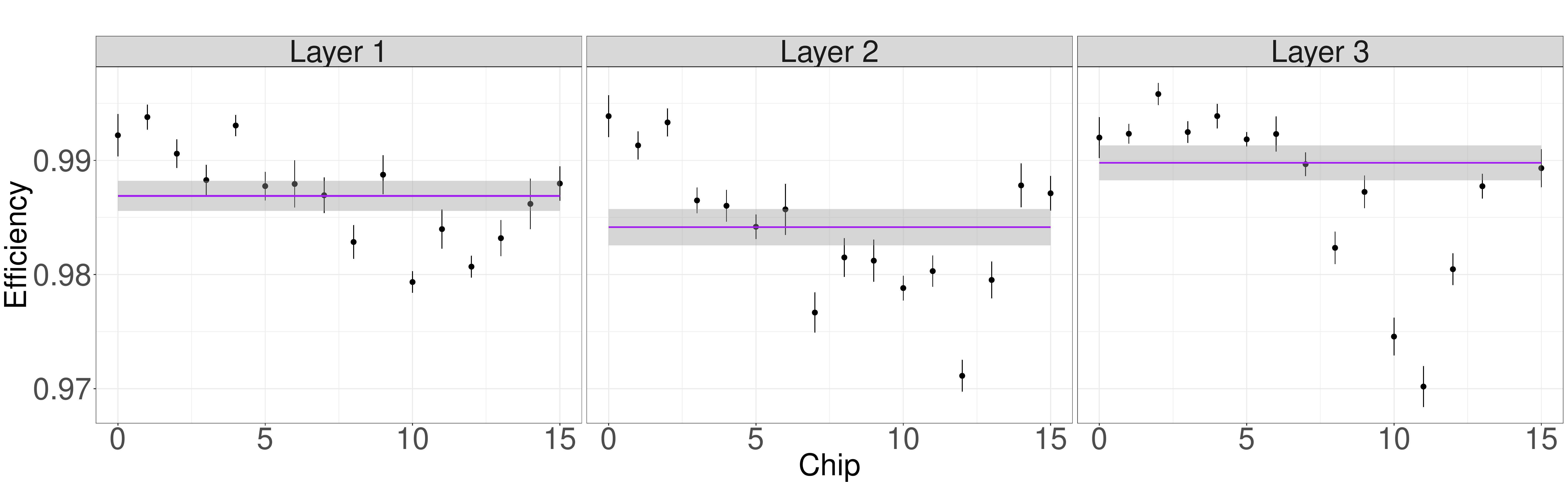}
\caption{\label{fig2}The efficiency averaged per chip excluding 2.9\% of
  outliers.}
\end{figure}

The resulting inefficiency is mainly due to the SKIROC trigger threshold (230
for this beam test). The probability that the pixel becomes efficient due to a
random noise is negligible. For 2.9\% of channels the inefficiency is found to
be $>20\%$. This is dominated by one chip in layer 3 (2.1\%). The rest is
dominated by the inefficiency due to masked pixels in the studied layer but
not in the trigger. The efficiency averaged over all channels without 2.9\% of
outliers per chip is 98--99\% and is shown in Figure~\ref{fig2}.

\begin{figure}[htbp]
\centering
\includegraphics[width=6.1cm]{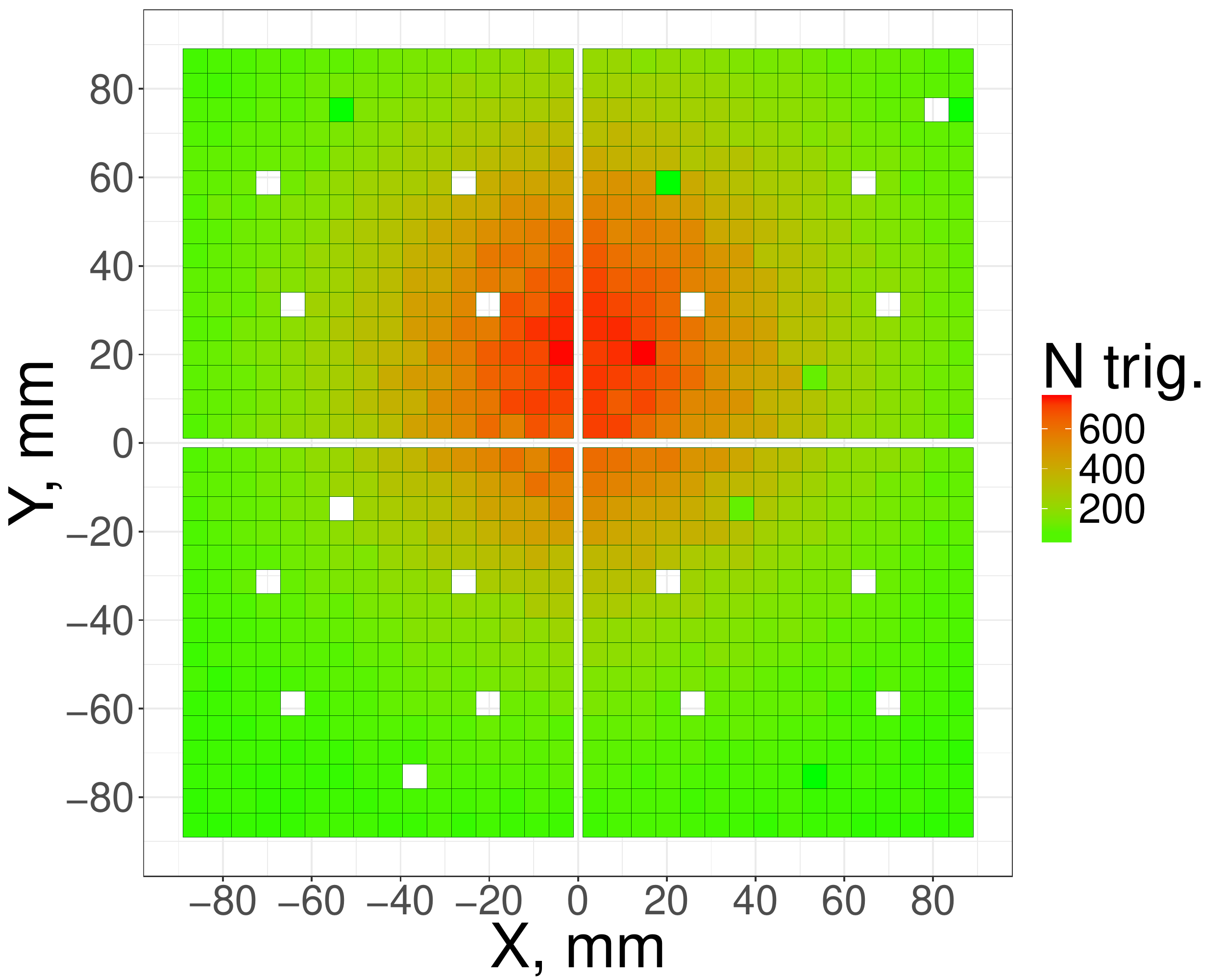}
\includegraphics[width=7.9cm]{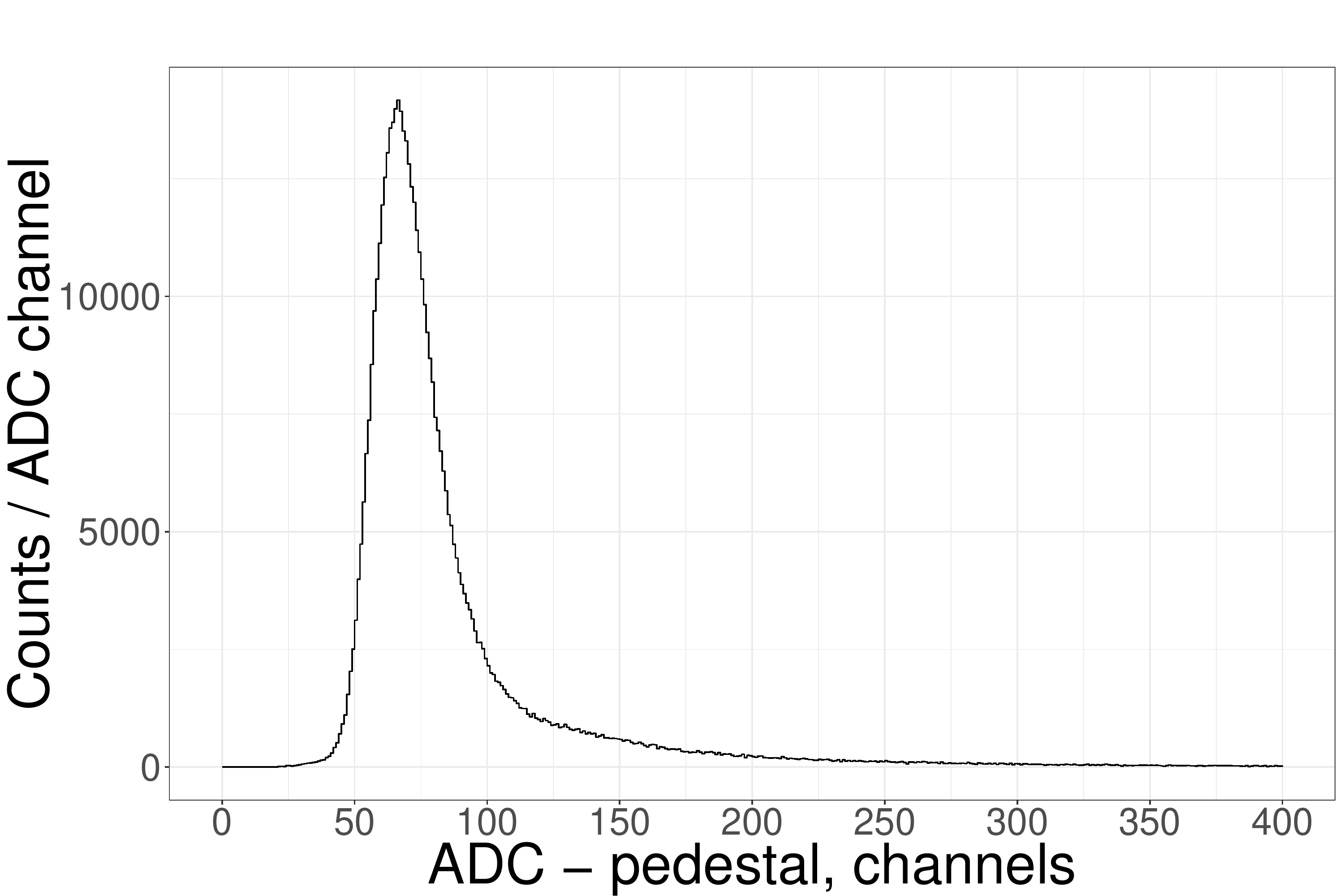}
\caption{\label{fig3}Left: first layer muon occupancy plot; right: 150~GeV
  triggered muon dE/dx energy deposition.}
\end{figure}

The transverse beam spot in the first layer and its muon ADC pedestal
subtracted spectrum summed over all channels after requiring the coincidence
with other two layers are shown in Figure~\ref{fig3}. The plots for the second
and the third layers are similar. The fraction of masked channels was 2.2\%.

One of the big advantages of the silicon detector is the gain uniformity
across the pixels.  This makes the calibration much easier and reduces the
systematics. The spread of the muon dE/dx most-probable-value across the
pixels before any calibration is measured to be 6.4\%. The most-probable-value
was calculated for simplicity as the truncated mean among 57\% of the lowest
pedestal subtracted ADC values in the region above 40. It was required that
100\% corresponds to at least 50 events, therefore due to a limited muon
statistics at the corners, only 83\% of all channels were calibrated in this
way.

\subsection{``Square'' events}
There is a special electrode at the silicon sensor periphery, called ``guard
ring'. It smooths the electric field at the edge and ensures low dark
currents. Due to technological reasons, in the current technological prototype
the guard ring potential is not fixed but floating. Therefore, a big energy
deposition there may fire many peripheral pixels via the capacitive coupling
and produce ``square''-like events. They have been observed in SiW ECAL
physical prototype. Contrary to the sensors of the physical prototype, in the
technological prototype their guard ring is segmented. This greatly reduces
the capacitive coupling.

The probability of ``square'' events was studied in the showers created by
positrons of maximal energy of 100 and 150~GeV traversing 8.4~radiation
lengths of the tungsten absorber in front of 3 ECAL layers. 100~GeV positrons
were shooting approximately at their centers, i.e. at four corners of the
silicon sensors. Therefore, the probability of the ``square'' events was
maximized. 150~GeV beam was shifted horizontally by 1 cm. About 90\% of the
transverse energy barycenters were contained in an ellipsoid with 3~cm
horizontal and 2~cm vertical axes. In these conditions at 150~GeV (100~GeV) we
observed 18 (69) ``square''-like patterns in one of 4 sensors out of 119~000
(390~000) events, which corresponds to the probability of $4\cdot10^{-5}$ per
sensor in both data samples.


\section{Conclusions}

The silicon detectors started to ``expand'' from the trackers to the
calorimeters in world-leading accelerator detectors.

The SiW ECAL technological prototype has reached the ILD design density of
channels and has demonstrated the ability to work in ILD ``power pulsing''
mode. It has MIP-to-noise ratio of $\sim$19 (with the optimal SKIROC gain, 5
times larger than in ILD), the muon efficiency of 98--99\% and the spread of
the MIP signals of 6.4\% measured in 3000 channels before any
calibration. With the improved design of the sensor guard rings the
probability of ``square'' events per sensor is measured to be well below
$10^{-4}$.  There is still a lot of work ahead, in particular, on improving
the trigger noises.

\acknowledgments

Supported by the H2020 project AIDA-2020, GA no. 654168.


\end{document}